\documentstyle[graphicx,amssymb,floats,prl,aps]{revtex}

\newcommand{\Msun}{{{\rm M}_\odot}}

\begin{document}
\draft
\twocolumn[\columnwidth\textwidth\csname@twocolumnfalse\endcsname
\title{Presupernova collapse models with improved weak-interaction rates}
\author{A. Heger$^1$, K. Langanke$^2$, G. Mart\'{\i}nez-Pinedo$^2$,
and S.E. Woosley$^1$}

\address{$^1$ UCO/Lick Observatory, University of California, Santa
  Cruz CA 95064, USA \\
  $^2$Institut for Fysik og Astronomi, {\AA}rhus Universitet, DK-8000
  {\AA}rhus C, Denmark}

\date{\today}
\maketitle

\begin{abstract}
Improved values for stellar weak interaction rates have been recently
calculated based upon a large shell model diagonalization.  Using
these new rates (for both beta decay and electron capture), we have
examined the presupernova evolution of massive stars in the range
$15$--$40\,\Msun$.  Comparing our new models with a standard set of
presupernova models by Woosley and Weaver, we find significantly
larger values for the electron-to-baryon ratio $Y_e$ at the onset of
collapse and iron core masses reduced by approximately $0.1\,\Msun$.
The inclusion of beta-decay accounts for roughly half of the
revisions, while the other half is a consequence of the improved
nuclear physics. These changes will have important consequences for
nucleosynthesis and the supernova explosion mechanism.
\end{abstract}

]

The late stages of massive stellar evolution are strongly influenced
by weak interactions which act to determine the core entropy and
electron-to-baryon ratio, $Y_e$, of the presupernova star.  Electron
capture reduces the number of electrons available for pressure
support, while beta decay acts in the opposite direction.  Both
processes generate neutrinos which, for densities $\rho \lesssim
10^{11}\,$g/cm$^3$, escape the star carrying away energy and entropy
from the core.

Electron capture and beta decay during the final evolution of a
massive star are dominated by Fermi and Gamow-Teller (GT) transitions.
While the treatment of Fermi transitions (important only in beta
decays) is straightforward, a correct description of the GT
transitions is a difficult problem in nuclear structure.  In their
pioneering work on the subject Fuller, Fowler and Newman
(FFN)~\cite{FFN} estimated electron-capture rates assuming a single GT
resonance. The properties of this resonance were derived on the basis
of the independent particle model, supplemented by Fermi contributions
and experimental data for low-lying transitions, whenever available.
These authors also noted the importance of the `backresonances' for
beta decay. These are excited states in the decaying nucleus which are
connected by strong GT transitions to low-lying states in the daughter
nucleus and, by thermal population and with increased phase space, can
significantly contribute to the stellar beta decay rates.

Recent experimental data shows that the GT distributions in nuclei are
quenched, compared to the independent particle model value, and
strongly fragmented over many states in the daughter nucleus. Both
effects are caused by residual interaction among the valence nucleons
and an accurate description of these correlations is essential for a
reliable evaluation of the stellar weak interaction rates due to the
strong phase space energy dependence, particularly of the stellar
electron capture rates.  The shell model is the only known tool to
reliably describe GT distributions in nuclei \cite{Wildenthal}. Its
application to iron mass nuclei in the middle of the $pf$-shell as
required in the presupernova collapse, however, has long been
inhibited due to the extremely large model space dimensions involved.
After significant progress in shell-model programming \cite{Caurier}
and hardware development the situation has changed very recently and
in Ref.~\cite{Caurier99} it has been demonstrated that
state-of-the-art diagonalization studies, typically involving a few 10
million configurations, are indeed able to reproduce all relevant
ingredients (GT$_\pm$ strength distributions for changing protons
(neutrons) into neutrons (protons), level spectra and half-lives) and
hence have the predictive power to reliably calculate stellar weak
interaction rates.  This program has recently been finished and
stellar weak-interaction rates for nuclei with $A=45$--65 have been
calculated based on the shell-model results, supplemented by
experimental data, wherever available. The shell-model rates have been
discussed and validated in \cite{Langanke00}.  It has been found that
for $pf$-shell nuclei the shell-model electron-capture rates are
smaller than the FFN rates by, on average, an order of magnitude, for
the reasons explained in \cite{Langanke00}.  The situation is
different for the beta decay as the shell model and FFN rates are of
the same magnitude for the most relevant nuclei to be identified
below.

To study the influence of the shell model rates on presupernova models
we have repeated the calculations of Woosley and Weaver
\cite{Weaver95} (henceforth WW) keeping all the stellar physics,
except for weak rates, as close to the original studies as possible.
The present calculations have incorporated the new shell-model weak
interaction rates (including electron capture, positron emission, and
beta decay) for nuclei with mass numbers $A=45$--65, supplemented by
rates from Oda {\it et al.}\cite{Oda} for lighter nuclei.  In
practice, the weak rates for these lighter nuclei were not very
important for determining the presupernova structure, but only
dominate prior to silicon burning. We also note that, for $sd$-shell
nuclei with $A=17-39$, the FFN rates agree rather well with the shell
model rates previously determined by Oda {\it et al.}\cite{Oda}.

The earlier calculations of WW, to which we compare, used the FFN
rates for electron capture and an older set of beta-decay rates taken
from \cite{Mazurek} and \cite{Hansen}.  Shortly after the models of WW
were calculated, it was recognized that these older beta-decay rates
were inadequate and that use of the larger values from FFN would
appreciably alter the results \cite{Fuller94}.  These are the first
models of the WW variety to incorporate realistic beta-decay rates and
electron-capture rates in a complete stellar model, even though beta
decays were included in some other models of massive stellar evolution
\cite{Nomoto}.  In a separate paper \cite{Heger00}, we will present
the comparison of models that use the full FFN rate set and the new
rate set, and in a future paper we will examine the evolution of stars
of other mass and metallicity.

\begin{figure*}[ht]
  \begin{center} \leavevmode
    \includegraphics[width=0.7\textwidth]{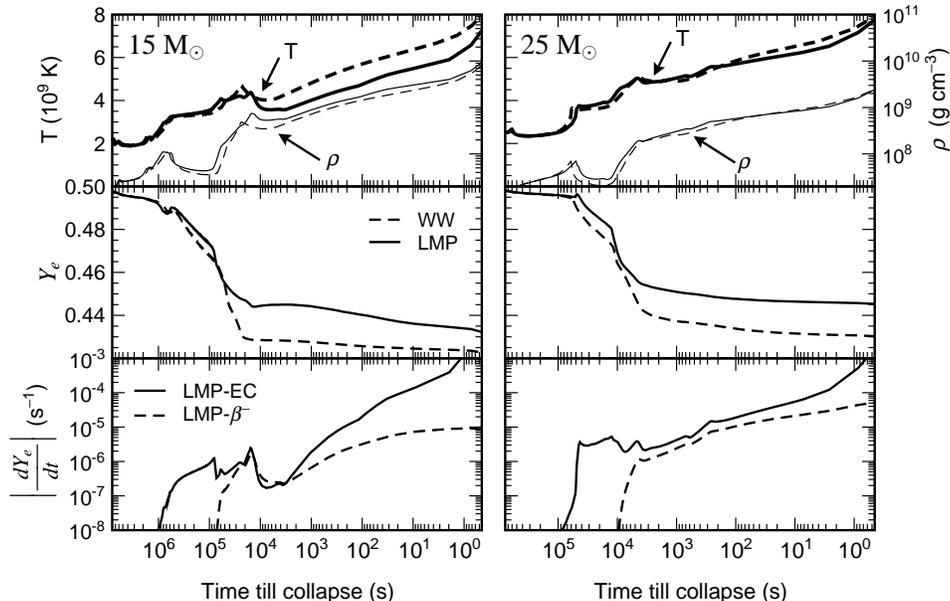}
    \caption{Comparison of the time evolution of key
      quantities at the center of a $15\,\Msun$ (left) and $25\,\Msun$
      (right) star between the models by Woosley \& Weaver models (WW,
      dashed)~\protect\cite{Weaver95} and the present ones using the
      shell model weak interaction rates (LMP, full lines).  The upper
      panels show the temperature (thick lines) and density (thin
      lines), the middle panels show the electron-to-baryon ratio,
      $Y_e$, and the lower panels show the time-derivative of $Y_e$
      due to electron capture and beta decay for the models using the
      LMP rates.  As reference points, in the $15\,\Msun$
      ($25\,\Msun$) star central silicon burning ignites at $7 \times
      10^5\,$s ($5.6 \times 10^4\,$s), terminates at $6.6 \times
      10^4\,$s ($10^4\,$s), silicon shell burning starts at $3.2
      \times 10^4\,$s ($4.7 \times 10^3\,$s) and core contraction set
      in at $t\sim280\,$s ($\sim230\,$s) before core collapse.  Note
      that the $dY_e/dt$ shown in the lower panels includes only the
      contribution of weak interaction processes to the total change
      of the central $Y_e$; during central silicon burning convection
      mixes down matter with higher $Y_e$ from layers
      above~\protect\cite{Heger00}.\label{fig1}}
    \end{center}
\end{figure*}

Fig.~\ref{fig1} shows the late evolution, following core oxygen
burning, of the central temperature and density in $15\,\Msun$ and
$25\,\Msun$ stars as well as the central value of the
electron-to-baryon ratio, $Y_e$. Time is measured here backwards from
the time of iron core collapse, which is arbitrarily t = 0.  Silicon
burning ignites with a mild ``flash'' and the core becomes convective
and expands.  For the $25\,\Msun$ star the temperature and density
trajectories in the new and old models are similar, but the
calculations with shell model rates have significantly larger values
of $Y_e$. This difference persists throughout the iron core, not just
at its center. Larger values of $Y_e$ also result in the $15\,\Msun$
star, but there even the density and temperature structures of the
presupernova star are appreciably altered.

To understand the origin of these differences, we will now explore the
role of the weak-interaction rates in greater detail. Weak processes
become particularly important in reducing $Y_e$ below 0.50 after
oxygen depletion ($\sim 10^7$~s and $10^6$~s before core collapse for
the $15\,\Msun$ and $25\,\Msun$ stars, respectively) and $Y_e$ begins
a decline that becomes precipitous during silicon burning.  Initially
electron capture occurs much more rapidly than beta decay. Since the
shell model rates are generally smaller than the FFN electron capture
rates, the initial reduction of $Y_e$ is smaller in the new models.
The reduction is less pronounced during silicon shell burning both
because the evolution time scale becomes quite short and because
nuclei near the valley of beta-stability, with smaller
weak-interaction rates, have already been produced (from a composition
that initially had $N=Z$). During this period the core matter is
composed of nuclei with $A<65$ which are carried in the calculation.

An important feature of the new models is that beta decay becomes
temporarily competitive with electron capture after silicon depletion
in the core and during silicon shell burning. That this would occur
was foreseen by \cite{Fuller94} on the basis of one-zone models.  Here
we see it occurring in a complete stellar model.  Moreover, the new
electron capture rates are smaller than FFN and thus offer less
resistance to beta decay. Dynamic weak equilibrium, in the sense
described by \cite{Fuller94}, thus occurs at larger values of $Y_e$.
Interestingly, by the time the iron core is actually collapsing, weak
equilibrium no longer exists. The increase in density closes the phase
space for beta decay and electron capture again predominates by a
large factor. It is this special characteristic of ``presupernova
models'' that lead some researchers in the past to miss the importance
of beta decay during a transient stage an hour or so {\em prior to
  collapse}.

While dynamic weak equilibrium is achieved in the $15\,\Msun$ model,
with the new rates, it is not in the $25\,\Msun$, though beta-decay
still offers a non-negligible resistance to electron capture even
there. This is in part due to the shorter time scale of silicon shell
burning in the more massive star and also the larger value of $Y_e$ in
the cores of stars with higher entropies (Fig.~\ref{fig1}).

The presence of an important beta-decay contribution has two effects.
Obviously it counteracts the reduction of $Y_e$ in the core, but
equally important, beta decays are an additional neutrino source and
thus add to the cooling of the core and a reduction in entropy
\cite{Fuller94}. This cooling can be quite efficient, as often the
average neutrino energy in the involved beta decays is larger than for
the competing electron captures. As a consequence the new models have
significant lower core temperatures than the WW models after silicon
burning, which is particularly pronounced for the $15\,\Msun$
star. During the contraction stage electron capture is again more
important than beta decays, associated with the increased electron
Fermi energy. Although the shell model rates are individually smaller
than the FFN electron capture rates, the effective electron capture
rate is larger in the new models as the evolution now proceeds along a
trajectory with larger $Y_e$ values involving nuclei with smaller
Q-values thus making electron capture energetically easier.

\begin{figure}[ht]
  \begin{center}
    \leavevmode
    \includegraphics[width=0.7\columnwidth]{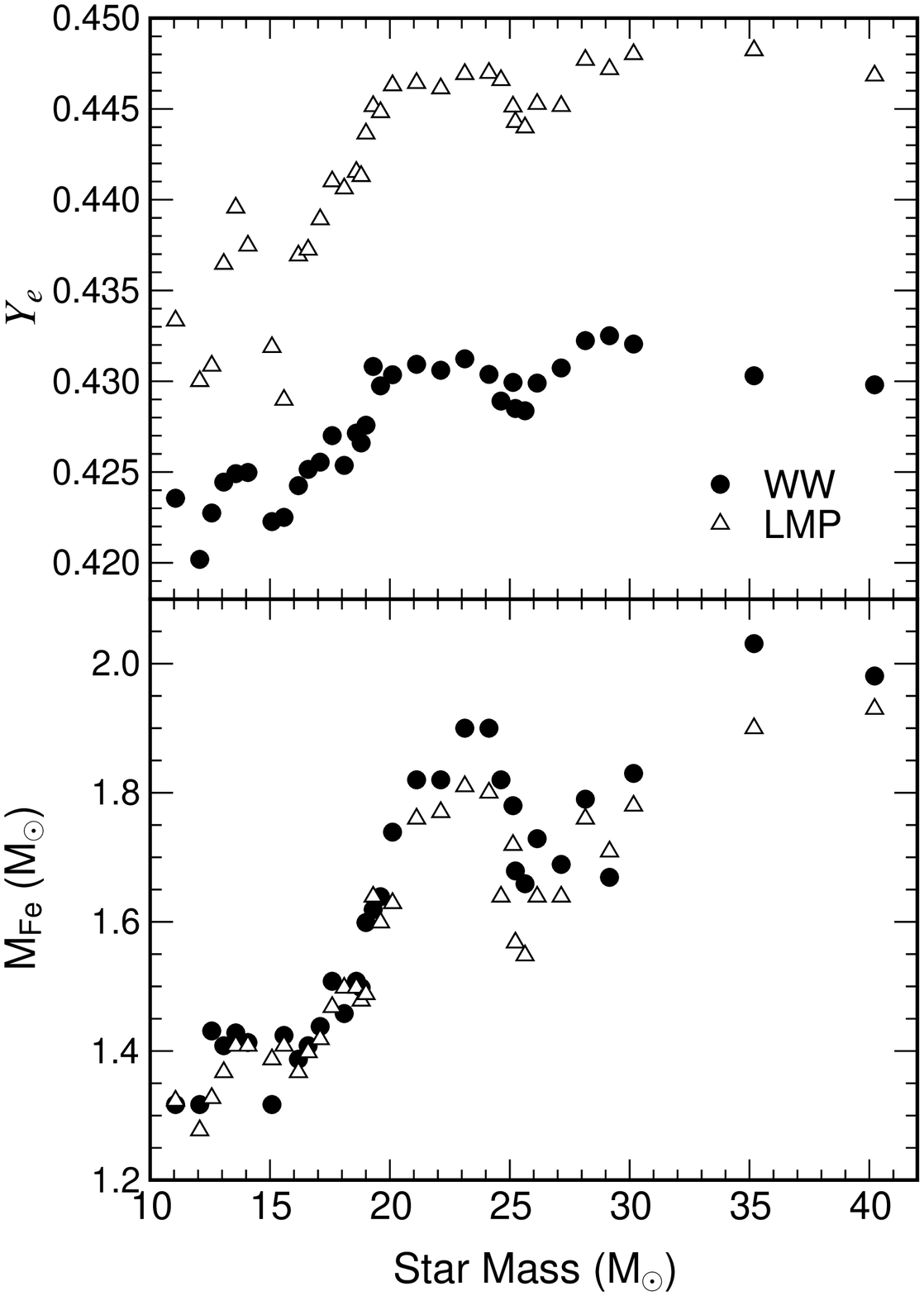}
    \caption{Comparison of the center values of $Y_e$ (upper part) and
      the iron core sizes (lower part) for 11--40~$\Msun$ stars
      between the WW models and the present ones using the LMP weak
      interaction rates.  }
    \label{fig2}    
  \end{center}
\end{figure}

In summary, the shell model weak interaction rates result in
significantly larger $Y_e$ values during the presupernova evolution of
a 15 and 25~$\Msun$ star than calculated by WW. Part - about half - of
the change is due to including beta-decay and the other half is due to
slower rates for electron capture.  Fig.~\ref{fig2} shows that this is
a general finding for all stars in the mass range $11$--$40\,\Msun$.
The central values of $Y_e$ at the onset of core collapse are
increased by $0.01$--$0.015$.  This is a significant effect.  We note
that the new models also result in lower core entropies for stars with
$M \lesssim 20\,\Msun$. The larger core $Y_e$ values and the lower
entropies suggest that these stars will have a larger homologous core
than currently assumed. For $M \gtrsim 20\,\Msun$, the new models
actually have a slightly larger entropy. Thus one might expect that 
increased electron capture rate on (the more abundant) free protons
will partly counteract the increase in $Y_e$ values during the subsequent
core collapse phase.  In general, core collapse calculations with
detailed neutrino transport are required before definite conclusions
about the explosion mechanism can be drawn. We will provide our models
to those wishing to attempt such calculations.

Another important property for the core collapse is the size of the
iron core, which we define as the mass interior to the point where the
composition becomes at least $50\,\%$ of iron group elements $(A \geq
48$). As is shown in Fig.~\ref{fig2}, the iron core masses are
generally smaller in the new models. The effect is larger for stars
more massive than $20\,\Msun$, while for the most common supernovae
(\mbox{$M \lesssim 20\,\Msun$}) the reduction is about $0.05\,\Msun$.
This reduction appears to be counterintuitive at first glance with
respect to the slower electron capture rates in the new models. It is,
however, related to changes in the entropy profile during silicon
shell burning which reduces the growth of the iron core just prior to
collapse~\cite{Heger00}. Clearly, though we have, for reasons of space
concentrated upon central values, it is the entire distribution of
entropy and $Y_e$ in the stellar interior that determines its final
evolution.

\begin{table*}[htb]
  \caption{Most important nuclei for electron capture and beta decay
    at selected points (characterized by temperature $T$, density
    $\rho$ and electron-to-baryon ratio $Y_e$) during the final 
    evolution of a 15~$\Msun$ and $25\,\Msun$ stars. The total
    electron capture $\lambda_{\text{ec}}$ and beta decay
    $\lambda_{\beta^-}$ rates are listed as well as the 3 dominating
    nuclei; the number in parentheses defines their percentage to the
    respective rates.} 
  \label{tab1}
  \renewcommand{\arraystretch}{1.1}
  \begin{tabular}{ccccccccccc}
    $T$ (K) & $\rho$ (g cm$^{-3}$) & $Y_e$ & $\lambda_{\text{ec}}$
    (s$^{-1}$) &  \multicolumn{3}{c}{electron capture} &
    $\lambda_{\beta^-}$ (s$^{-1}$) &
    \multicolumn{3}{c}{beta decay} \\ \hline 
    \multicolumn{11}{c}{15 $\Msun$}\\
    \hline      
    $3.39 \times 10^9$ & $4.50 \times 10^7$ & 0.480 & $5.17 \times 10^{-7}$
    & $^{54}$Fe (29) & $^{55}$Fe (25) & $^{53}$Mn (11) & $6.08 \times
    10^{-11}$ & $^{54}$Mn (67) & $^{55}$Mn (8) & $^{32}$P (7) \\ 
    $3.82 \times 10^9$ & $7.26 \times 10^7$ & 0.464 & $3.30 \times
    10^{-7}$ & $^{55}$Fe (41) & $^{57}$Co (10) & $^{53}$Mn (9) & $6.73
    \times 10^{-9}$ & $^{56}$Mn (45) & $^{60}$Co (18) & $^{55}$Mn (11) \\
    $4.13 \times 10^9$ & $2.89 \times 10^8$ & 0.450 & $6.86 \times
    10^{-7}$ & $^{57}$Fe (54) & $^{61}$Ni (21) & $^{56}$Fe (14) &
    $4.10 \times 10^{-7}$ & $^{56}$Mn (36) & $^{52}$V (12) & $^{57}$Mn
    (10) \\
    $4.41 \times 10^9$ & $1.30 \times 10^9$ & 0.442 & $7.57\times
    10^{-6}$ & $^{57}$Fe (22) & $^{53}$Cr (14) & $^{55}$Mn (13) &
    $1.74 \times 10^{-6}$ & $^{58}$Mn (34) & $^{62}$Co (17) &
    $^{64}$Co (12) \\
    $7.25 \times 10^9$ & $9.36 \times 10^9$ & 0.432 & $9.21 \times
    10^{-3}$ & $^{65}$Ni (14) & $^{59}$Fe (7) & $^{52}$V (7) & $8.45
    \times 10^{-6}$ & $^{64}$Co (22) & $^{58}$Mn (19) & $^{54}$V (13)
    \\\hline
    \multicolumn{11}{c}{25 $\Msun$}\\
    \hline      
    $3.79\times 10^9$ & $2.89\times 10^7$ & 0.487 & $3.18 \times
    10^{-6}$ & $^{53}$Fe (23) & $^{55}$Co (20) & $^{56}$Ni (19) &
    $1.53\times 10^{-11}$ & $^{54}$Mn (49) & $^{55}$Fe (17) &
    $^{58}$Co (9) \\
    $4.17\times 10^9$ & $3.71\times 10^7$ & 0.476 & $4.23\times
    10^{-6}$ & $^{54}$Fe (21) & $^{55}$Co (14) & $^{55}$Fe (11) &
    $8.12\times 10^{-10}$ & $^{54}$Mn (37) & $^{58}$Co (30) &
    $^{55}$Fe (8) \\ 
    $5.03\times 10^9$ & $1.82\times 10^8$ & 0.456 & $3.84\times
    10^{-6}$ & $^{56}$Fe (17) & $^{55}$Fe (13) & $^{61}$Ni (10) &
    $1.00\times 10^{-6}$ & $^{56}$Mn (45) & $^{52}$V (13) & $^{60}$Co
    (10) \\
    $5.57\times 10^9$ & $5.05\times 10^8$ & 0.449 & $1.45\times
    10^{-5}$ & $^{57}$Fe (16) & $^{56}$Fe (11) & $^{53}$Cr (9) &
    $7.61\times 10^{-6}$ & $^{56}$Mn (19) & $^{58}$Mn (14) & $^{55}$Cr
    (10) \\
    $7.75\times 10^9$ & $2.42\times 10^9$ & 0.445 & $1.95\times
    10^{-3}$ & $^1$H (32) & $^{53}$Cr (9) & $^{57}$Fe (7) &
    $5.17\times 10^{-5}$ & $^{58}$Mn (18) & $^{55}$Cr (13) & $^{57}$Mn
    (7) \\
\end{tabular}
\end{table*}

Confidence in the shell model rates stems from the recent measurements
of GT distributions in iron mass nuclei which are all well reproduced
by the shell model calculations~\cite{Caurier99}. However, the energy
resolutions of these pioneering (n,p) charge-exchange studies
performed at \mbox{TRIUMF} has been rather limited ($\sim
1$--$1.5\,$MeV) and they have been performed for stable nuclei only.
These limitations are likely to be overcome in the near future as
measurements with charge-exchange reactions like $($d$,{}^2$He),
$($t$,{}^3$He) promise data with an order of magnitude improved
resolution. Furthermore, after radioactive ion beam facilities will
become operational it will be possible by inverse techniques also to
determine the GT distributions for unstable nuclei.  Of course,
laboratory experiments cannot measure directly the relevant stellar
rates as these involve, for example, electron capture or beta decays
from excited states. Nevertheless, high-resolution charge-exchange or
beta-decay experiments are important for two reasons: First, they are
stringent constraints for the nuclear models and their predictive
powers. Second, such experiments can determine the energy positions of
the daughter states for the GT transitions which can then be used
directly in the determination of the rates.

To guide such experiments we have attempted to identify the most
important nuclei for electron-capture and beta decay during the final
stages of stellar evolution.  The relevant quantity is the product of
abundance of the nuclear species in the core composition and electron
capture rate (or beta decay rate).  Table~\ref{tab1} lists the most
important nuclei at selected points during the final evolution of our
$15\,\Msun$ and $25\,\Msun$ stars. Since the $Y_e$ values are
significantly larger in the new models, the important flows now
involve nuclei much closer to stability so that several of the most
important electron capturing nuclei (e.g., $^{54,56,58}$Fe, $^{55}$Mn,
$^{53}$Cr) are stable.  Beta decay, on the other hand, mainly involves
unstable manganese and cobalt isotopes. However, the lifetimes are
long enough to allow for an experimental determination of the relevant
GT strength distribution once radioactive ion beam facilities become
operational.  We finally mention that the backresonances contribute
noticeably to the stellar beta decay rates for these isotopes making
also measurement of the GT$_+$ strength on the daughter nuclei (e.g.
iron and nickel isotopes) very useful.

We would like to thank G. M. Fuller for helpful discussions.  This
work has been partly supported by the Danish Research Council, the
Carlsberg Foundation, by NATO grant CRPG973035, by the National
Science Foundation (NSF-AST-9731569), by the US Department of Energy
(DOE/LLNL B347885), and by the Alexander von Humboldt-Stiftung
(FLF-1065004).


\begin{thebibliography}{99}
  
\bibitem{FFN} G.M. Fuller, W.A. Fowler and M.J. Newman, Astrophys. J.
  Suppl. Ser. {\bf 42}, 447 (1980); Astrophys. J. Suppl.  Ser. {\bf
    48}, 279 (1982); Astrophys. J. {\bf 252}, 715 (1982)
  
\bibitem{Wildenthal} B.A. Brown and B.H. Wildenthal, Ann. Rev. Nucl.
  Part.  Sci. {\bf 38}, 29 (1988)
  
\bibitem{Caurier} E. Caurier, computer code {\sc antoine\/}, IReS
  Strasbourg (1989)
  
\bibitem{Caurier99} E. Caurier, K. Langanke, G. Mart\'{\i}nez-Pinedo
  and F. Nowacki, Nucl. Phys. A {\bf 653}, 439 (1999)
  
\bibitem{Langanke00} K. Langanke and G. Mart\'{\i}nez-Pinedo, Nucl.
  Phys. A {\bf 673}, 481 (2000)
  
\bibitem{Weaver95} S. E. Woosley and T. A. Weaver, Astrophys. J.
  Suppl. Ser. {\bf 101}, 181 (1995).
  
\bibitem{Oda} T. Oda, M. Hino, K. Muto, M. Takahara and K. Sato, At.
  Data Nucl. Data Tables {\bf 56}, 231 (1994)
  
\bibitem{Mazurek} T. Mazurek, PhD Thesis, Yeshiva Univ., (1973)

\bibitem{Hansen} C. J. Hansen, PhD Thesis, Yale Univ. (1966)

\bibitem{Fuller94} M. B. Aufderheide, I. Fushiki, G. M. Fuller and T.
  A. Weaver, Apstrophys. J. {\bf 424}, 257 (1994). 
  
\bibitem{Nomoto} F.-K. Thielemann, K. Nomoto, and M. Hashimoto,
 Astrophys. J., {\bf 460}, 408, (1996)

\bibitem{Heger00} A. Heger, S.E. Woosley, G. Mart\'{\i}nez-Pinedo and
  K.  Langanke, in preparation
  
\end{thebibliography}
\end{document}